\documentstyle[psfig,eqsecnum,aps]{revtex}
\begin{document}
\draft
\title{A deformed QRPA formalism for single and two-neutrino double 
beta decay}

\author{R. \'Alvarez-Rodr\'{\i}guez, P. Sarriguren, E. Moya de Guerra}
\address{Instituto de Estructura de la Materia,
Consejo Superior de Investigaciones Cient\'{\i }ficas, \\
Serrano 123, E-28006 Madrid, Spain}
\author{L. Pacearescu, Amand Faessler}
\address{Institut f\"ur Theoretische Physik, Universit\"at T\"ubingen, 
D-72076 T\"ubingen, Germany}
\author{F. \v Simkovic}
\address{Department of Nuclear Physics, Comenius University,
SK-842 15, Bratislava, Slovakia}
\date{\today}
\maketitle

\begin{abstract}

We use a deformed QRPA formalism to describe simultaneously the  
energy distributions of the single beta Gamow-Teller strength and
the two-neutrino double beta decay matrix elements. Calculations 
are performed in a series of double beta decay partners with 
A = 48, 76, 82, 96, 100, 116, 128, 130, 136 and 150, using deformed
Woods-Saxon potentials and deformed Skyrme Hartree-Fock mean fields.
The formalism includes a quasiparticle deformed basis and residual 
spin-isospin forces in the particle-hole and particle-particle channels. 
We discuss the sensitivity of the parent and daughter Gamow-Teller 
strength distributions in single beta decay, as well as the sensitivity 
of the double beta decay matrix elements to the deformed mean field 
and to the residual interactions. Nuclear deformation is found to be
a mechanism of suppression of the two-neutrino double beta decay. 
The double beta decay matrix elements are found to have maximum values 
for about equal deformations of parent and daughter nuclei. They 
decrease rapidly when differences in deformations increase. 
We remark the importance of a proper simultaneous description of
both double beta decay and single Gamow-Teller strength distributions.
Finally, we conclude that for further progress in the field it would 
be useful to improve and complete the experimental information on the 
studied Gamow-Teller strengths and nuclear deformations.
\end{abstract}

\pacs{PACS:  21.60.Jz, 23.40.Hc, 23.40.Bw}

\section{Introduction}

The recent experimental confirmation of neutrino oscillations has
reinforced the interest of nuclear processes involving neutrinos, 
see \cite{bernabeu} and references therein. Nuclear double beta decay 
is a rare second order weak interaction process that takes place when 
the transition to the intermediate nucleus is energetically forbidden 
or highly retarded. Two main decay modes are expected in this process. 
The two neutrino mode, involving the emission of two electrons and 
two neutrinos, and the neutrinoless mode with no neutrino leaving the 
nucleus. While the first type of process is perfectly compatible with 
the Standard Model, the second one violates lepton number conservation 
and its observation is linked to the existence of a massive Majorana 
neutrino. For this reason, considerable experimental and theoretical 
effort is being devoted to the study of this process \cite{2breview}.

From the theoretical point of view, one particular source of uncertainty 
is the evaluation of the nuclear matrix elements involved in the process. 
They have to be calculated as accurately as possible to obtain reliable 
estimates for the limits of the double beta decay half-lives. Since 
the nuclear wave functions and the underlying theory for treating the 
neutrinoless and the two-neutrino modes are similar, the usual procedure 
is to test first the nuclear structure component of the two-neutrino 
mode against the available experimental information on half-lives.

The proton-neutron quasiparticle random phase approximation (pnQRPA or 
QRPA in short) is one of the most reliable and widely used microscopic 
approximations for calculating the correlated wave functions involved 
in $\beta$ and double beta decay processes. The method was first 
studied in Ref. \cite{halb} to describe the $\beta$ strength and was 
also successfully applied to the description of double beta decay 
\cite{2bqrpa} after the inclusion of a particle-particle ($pp$) 
residual interaction, in addition to the particle-hole ($ph$) usual 
channel. Many more extensions of the QRPA method have been proposed in 
the literature, see Ref. \cite{extension} and references therein.

An extension of the pnQRPA method to deal with deformed nuclei was done 
in Ref. \cite{kru}, where a Nilsson potential was used to generate single 
particle orbitals. Subsequent extensions including Woods-Saxon type 
potentials \cite{moll}, residual interactions in the particle-particle 
channel \cite{homma}, selfconsistent deformed Hartree-Fock mean fields 
with consistent residual interactions \cite{sarr} and selfconsistent 
approaches in spherical neutron-rich nuclei \cite{doba}, can also be 
found in the literature. In Refs. \cite{sarr,gese}, $\beta $-decay 
properties were studied on the basis of a deformed selfconsistent 
HF+BCS+QRPA calculation with density dependent effective interactions 
of Skyrme type. A deformed QRPA approach based on a phenomenological 
deformed Woods-Saxon potential was used to calculate the Gamow-Teller 
strength distributions for the two decay branches in double beta decay 
of $^{76}$Ge \cite{gese,doublege}.

Recently, the issue of nuclear deformation, which has usually been 
ignored in the QRPA-like treatments of nearly spherical nuclei, was 
raised in Ref. \cite{doublege,feni}. In Ref. \cite{doublege} it was 
found that differences in deformation between the initial and final 
nuclei can have large effects on the  double beta decay half-lives. 
Within the deformed QRPA using axially-symmetric symmetric
Woods-Saxon single particle basis the particular case of the 
two-neutrino double beta decay ($2\nu\beta\beta$-decay) of $^{76}$Ge 
was analyzed \cite{doublege}. The effect of deformation on the other
double beta decay processes of experimental interest has not been 
sufficiently studied yet, see \cite{doublege,raduta,radescu} and 
references therein. 

In this work we extend these deformed calculations \cite{sarr,doublege}
by studying first the single $\beta$ branches that build up the double 
beta process and then the $2\nu\beta\beta$-decay process itself. 
We focus on the $\beta^-$ Gamow-Teller (GT) transitions of the double 
beta emitters as well as on the $\beta^+$ Gamow-Teller transitions of 
the daughter nuclei ending up to the same intermediate virtual nucleus. 
The cases considered are those where the two-neutrino double beta 
decay half-lives have been measured, namely
\begin{eqnarray*}
&^{48}Ca\rightarrow ^{48}Ti;\qquad &^{76}Ge\rightarrow ^{76}Se; \\
&^{82}Se\rightarrow ^{82}Kr;\qquad &^{96}Zr\rightarrow ^{96}Mo; \\
&^{100}Mo\rightarrow ^{100}Ru;\qquad &^{116}Cd\rightarrow ^{116}Sn; \\
&^{128}Te\rightarrow ^{128}Xe;\qquad &^{130}Te\rightarrow ^{130}Xe; \\
&^{136}Xe\rightarrow ^{136}Ba;\qquad &^{150}Nd\rightarrow ^{150}Sm.
\end{eqnarray*}

In sect. II, we present a brief summary containing the basic points in 
our theoretical description. Section III contains the results obtained 
for the ground state properties of the nuclei mentioned above. In sect. 
IV we present our results for the GT strength distributions and discuss 
their dependence on the deformed mean field and residual interactions. 
Sect. V contains the results for the two-neutrino double beta decay
calculations. The conclusions are given in sect. VI.

\section{Brief description of the theory}

In this Section we summarize briefly the theoretical formalism used to 
describe the Gamow-Teller transitions. More details can be found in 
Refs. \cite{sarr,doublege,gese}.

The single particle energies and wave functions are generated from two
different methods to construct the deformed mean field, which is 
assumed to be axially symmetric. In one case we start from a deformed 
Woods-Saxon potential. The parameters of this potential are taken from 
Ref.\cite{tanaka}. The isospin dependence of this parametrization 
allows one to extend it to any mass region. Previous QRPA calculations 
have shown that it provides a realistic description of the ground state 
properties of deformed nuclei as well as good results on $M1$ 
excitations \cite{m1} for nuclei in various mass regions. The quadrupole 
deformation of the WS potential is determined by fitting the 
microscopically calculated quadrupole moment to the corresponding 
experimental value. The hexadecapole deformation is expected to be 
small for these nuclei and we assume it is equal to zero. 

In the other method we follow a selfconsistent Hartree-Fock procedure
to generate microscopically the deformed mean field. This is done with
density dependent effective interactions of Skyrme type. Contrary to the
previous case, the equilibrium deformation of the nucleus is obtained now
selfconsistently as the shape that minimizes the energy of the nucleus.
In this work we present the results obtained with the most common of the
Skyrme forces, Sk3 \cite{beiner} although sometimes we also show for 
comparison results obtained with the force SG2 \cite{giai}.

In both schemes, WS and HF, the single-particle wave functions are 
expanded in terms of the eigenstates of an axially symmetric harmonic 
oscillator in cylindrical coordinates, using eleven major shells in the 
expansion. Pairing correlations between like nucleons are included 
similarly in both cases in the BCS approximation with fixed gap parameters 
for protons and neutrons. The gap parameters are determined 
phenomenologically from the odd-even mass differences through a symmetric 
five term formula involving the experimental binding energies. The values 
obtained from this procedure for the nuclei under consideration can be 
seen in Table 1.

The deformed quasiparticle mean field is now complemented with a 
spin-isospin separable residual interaction that contains two parts, an 
attractive particle-hole and a repulsive particle-particle. The coupling 
strengths of these forces $\chi ^{ph}_{GT}$ and $\kappa ^{pp}_{GT}$ are 
defined as positive. The strength of the $ph$ force is usually determined 
by adjusting the calculated positions of the GT giant resonances to 
experiment. The strength of the $pp$ force is determined by fitting the 
$\beta$-decay half-lives of $\beta$ emitters. This fitting procedure was 
systematically carried out in Ref. \cite{homma}, where the strengths 
$\chi ^{ph}_{GT}$, and $\kappa ^{pp}_{GT}$ were considered to be smooth 
functions of the mass number $A$. The result found using a Nilsson 
potential as the deformed mean field was 
$\chi ^{ph}_{GT}= 5.2 \; /A^{0.7}$ MeV and 
$\kappa ^{pp}_{GT}= 0.58\; /A^{0.7}$ MeV. Nevertheless, this 
parametrization clearly depends on the  model used for single particle 
wave functions and energies, as well as on the set of experimental data 
considered. Therefore, these coupling strengths can be used as a 
reference but cannot be safely extrapolated to different mean fields or 
different mass regions. As we shall see in the next section, the 
strengths from Ref. \cite{homma} reproduce well the data when using the 
WS potential, but one needs a somewhat smaller value of $\chi ^{ph}_{GT}$ 
to reproduce the GT resonance with the HF mean field.

We introduce the proton-neutron QRPA phonon operator for GT excitations 
in even-even nuclei

\begin{equation}
\Gamma _{\omega _{K}}^{+}=\sum_{\pi\nu}\left[ X_{\pi\nu}^{\omega _{K}}
\alpha _{\nu}^{+}\alpha _{\bar{\pi}}^{+}+Y_{\pi\nu}^{\omega _{K}}
\alpha _{\bar{\nu}} \alpha _{\pi}\right]\, ,  \label{phon}
\end{equation}
where $\alpha ^{+}\left( \alpha \right) $ are quasiparticle creation
(annihilation) operators, $\omega _{K}$ are the RPA excitation energies, 
and $X_{\pi\nu}^{\omega _{K}},Y_{\pi\nu}^{\omega _{K}}$ the forward and 
backward amplitudes, respectively. The solution of the QRPA equations 
can be found solving first a dispersion relation \cite{hir}, which is 
of fourth order in the excitation  energies $\omega_K$. 

In the intrinsic frame the GT transition amplitudes connecting the QRPA 
ground state $\left| 0\right\rangle$ to one phonon states 
$\left| \omega _K \right\rangle$ satisfying

\begin{equation}
\Gamma _{\omega _{K}} \left| 0 \right\rangle =0 \qquad
\Gamma ^+ _{\omega _{K}} \left| 0 \right\rangle = \left| 
\omega _K \right\rangle 
\end{equation}
are given by

\begin{equation}
\left\langle \omega _K | \sigma _K t^{\pm} | 0 \right\rangle = 
\mp M^{\omega _K}_\pm \, .
\end{equation}
where

\begin{equation}
M_{-}^{\omega _{K}}=\sum_{\pi\nu}\left( q_{\pi\nu}
X_{\pi\nu}^{\omega _{K}}+ \tilde{q}_{\pi\nu}
Y_{\pi\nu}^{\omega _{K}}\right) \, ; \qquad 
M_{+}^{\omega _{K}}=\sum_{\pi\nu}\left( \tilde{q}_{\pi\nu}
X_{\pi\nu}^{\omega _{K}}+
q_{\pi\nu}Y_{\pi\nu}^{\omega _{K}}\right) \, ,
\end{equation}
and
\begin{equation}
\tilde{q}_{\pi\nu}=u_{\nu}v_{\pi}\Sigma _{K}^{\nu\pi };\ \ \ 
q_{\pi\nu}=v_{\nu}u_{\pi}\Sigma _{K}^{\nu\pi}\, ;\qquad
\Sigma _{K}^{\nu\pi}=\left\langle \nu\left| \sigma _{K}\right| 
\pi\right\rangle \, ,
\label{qs}
\end{equation}
where $v'$s are occupation amplitudes ($u^2=1-v^2$).

Finally, the GT strength $B(GT^\pm)$ in the laboratory system for a 
transition $I_iK_i (0^+0) \rightarrow I_fK_f (1^+K)$ can be obtained 
as
\begin{equation}
B(GT^\pm ) = \frac{g_A^2}{4\pi}\left[ \delta_{K,0}\left\langle 
\omega_{K} \left| \sigma_0t^\pm \right| 0 \right\rangle ^2 +
2\delta_{K,1} \left\langle \omega_{K} \left| \sigma_1t^\pm \right| 
0 \right\rangle ^2 \right] .
\label{bgt}
\end{equation}

The $2\nu\beta\beta$ decay is described in second order perturbation
of the weak interaction as two successive Gamow-Teller transitions
via virtual intermediate $1^+$ states. The half-life of the
$2\nu\beta\beta$ decay
\begin{equation}
T^{2\nu}_{1/2}\,\left( 0^+_{\rm gs} \rightarrow 0^+_{\rm gs}
\right)  = \left[ G^{2\nu} \left| M^{2\nu}_{\rm GT} \right| ^2 
\right] ^{-1}
\label{t2nu}
\end{equation}
is given as a product of a phase space integral $G^{2\nu}$ and the 
Gamow-Teller transition matrix element $M^{2\nu}_{\rm GT} $, which 
contains the nuclear structure effects. For a transition connecting 
initial and final ground states, it is given by

\begin{equation}
 M^{2\nu}_{\rm GT}=\sum_{K}\sum_{m_i,m_f} 
\frac{\left\langle 0^+_f \left| \right| \sigma_K t^- \left| \right| 
\omega_K^{m_f}\right\rangle
\left\langle \omega_K^{m_f} \left| \right. \omega_K^{m_i} \right\rangle
\left\langle \omega_K^{m_i} \left| \right| \sigma_K t^- \left| \right| 
0^+_i \right\rangle }
{\left( \omega_K^{m_f}+\omega_K^{m_i}\right) /2 }\, ,
\label{m2nu}
\end{equation}
where $K=0,\pm 1$ and $m_i,m_f$ label the number of intermediate $1^+$ 
RPA states $\omega_K^{m_i},\, \omega_K^{m_f}$ reached from the initial 
$|0^+_i>$ and final $|0^+_f>$ nuclear ground states, respectively.
The overlap $<\omega_K^{m_f} | \omega_K^{m_i}>$ is needed to take into
account the non-orthogonality of the intermediate states reached from
the initial ground state to those reached from the final ground state. 
It is given by \cite{doublege}

\begin{equation}
\left\langle \omega_K^{m_f} | \omega_K^{m_i} \right\rangle =
\sum_{\ell_i\, \ell_f} \left[ X_{\ell_f}^{\omega_K^{m_f}} 
X_{\ell_i}^{\omega_K^{m_i}} - Y_{\ell_f}^{\omega_K^{m_f}}
Y_{\ell_i}^{\omega_K^{m_i}}\right] R_{\ell_f \ell_i}
\left\langle  BCS_f | BCS_i\right\rangle \, ,
\end{equation}
where $\ell_i\, \ell_f$ label the quasiparticle $\pi \nu$ pairs for 
the initial and final nucleus, respectively. The factor 
$ R_{\ell_f \ell_i}$ includes the overlap of single particle wave 
functions of the initial and final nuclei \cite{doublege}

\begin{equation}
R_{\ell \ell '} = <\pi |\pi '><\nu |\nu '> \left( u^i_\pi u^f_{\pi '}
+v^i_\pi v^f_{\pi '} \right) \left(  u^i_\nu u^f_{\nu '}
+v^i_\nu v^f_{\nu '} \right) . 
\end{equation}
The BCS overlap factor $<BCS_f | BCS_i>$ is derived in 
Ref. \cite{doublege}. An approximate expression is given by

\begin{equation}
\left\langle  BCS_f | BCS_i\right\rangle \approx
\prod_{\Omega_\pi}\prod _{k=1}^{N_{\Omega_\pi}} 
\left( u_k^f u_k^i + v_k^f v_k^i \right)
\prod_{\Omega_\nu}\prod _{j=1}^{N_{\Omega_\nu}} 
\left( u_j^f u_j^i + v_j^f v_j^i \right) \, ,
\label{bcsif}
\end{equation}
with $N_{\Omega}$ the number of single particle states with the same 
values of parity and projection $\Omega$ of the full angular momentum 
on the nuclear symmetry axis.

We note that for the case of the spherical QRPA the derivation of the 
overlap factor of the intermediate nuclear states generated from the 
initial and final nuclei was outlined in Ref. \cite{spf}. For the case 
of the deformed QRPA the generalization of this derivation was presented 
in \cite{doublege}. There, the importance of the BCS overlap factor, 
which is an integral part of the overlap factor of the two sets of the 
intermediate nucleus, on the evaluation of the double beta decay nuclear 
matrix elements was maintained. Then, this procedure of the derivation 
of the overlap factor of intermediate nuclear states was followed also
in Ref. \cite{radescu}, however,  with a significant approximation. The 
role of  the BCS overlap factor was neglected. Of course, this can affect 
the final results significantly, especially in the cases when 
deformations of the initial and final nuclei are different. 

\section{Ground state properties}

In this Section we present results for the bulk properties of the 
nuclei under study based on the quasiparticle mean field description. 
We consider both WS potential and Skyrme HF approaches.

In the case of HF, the first step is to study the energy surfaces as 
a function of deformation. For this purpose we perform constrained 
calculations \cite{constraint}, minimizing the HF energy under the 
constraint of keeping fixed the nuclear deformation. We can see in Fig. 1 
the total HF energy plotted versus the quadrupole deformation parameter
\begin{equation}
\beta = \sqrt{\frac{\pi}{5}} \frac{Q_p}{Zr_c^2} \, ,
\label{betadef}
\end{equation}
defined in terms of the microscopically calculated quadrupole moment 
$Q_p$ and charge root mean square radius $r_c$. 

The results in Fig. 1 correspond to HF calculations with the force Sk3, 
which are qualitatively similar to the results obtained with other 
Skyrme forces, such as SG2 and SLy4 \cite{sly4}. We observe that the 
HF calculation predicts in some instances (Ge, Se, Zr, Cd, Sn, Te, Xe) 
the existence of two energy minima close in energy, giving rise to 
possible shape isomers in these nuclei. Solid lines represent the energy 
curves of the parent nuclei suffering the double beta decay, while dashed 
lines represent the energy curves of the corresponding daughter nuclei.

We can see in Table 1 the experimental and the microscopically calculated 
charge root mean square radii $r_c$ with the forces Sk3 and SG2. We quote
the two, oblate-prolate, results in those cases where the energies for the
two shapes are very close. The values obtained for the charge radii are in 
good agreement with the experimental values from Ref. \cite{radiiexp}. They 
are also in good agreement with the results obtained from relativistic mean 
field calculations \cite{ring}.

In Table 2 we can see the theoretical and experimental quadrupole 
deformation. Experimental values have been extracted from the measured 
quadrupole moments from two different methods. In the first one the 
quadrupole deformation is obtained from Eq.(\ref{betadef}), using the 
empirical intrinsic moments derived from the laboratory moments of 
Ref.\cite{raghavan} assuming a well defined deformation. In the second 
case the quadrupole deformations are taken from Ref. \cite{raman}, where 
they were derived from experimental values of $B(E2)$ strengths. In this 
case the sign cannot be extracted.

Our theoretical values have been derived microscopically from the forces 
Sk3 and SG2, using the intrinsic quadrupole moments obtained as ground 
state expectations of the $Q_{20}$ operator and the microscopic charge 
radii quoted in Table 1. We show the results obtained for the equilibrium 
shapes using the Skyrme forces Sk3 and SG2. In those cases where a second 
minimum appears at close energy, we also show within square brackets the
corresponding deformation. We compare our results with the results from 
relativistic mean field calculations of Ref. \cite{ring} and with results 
from systematic calculations \cite{moeller} based on macroscopic-microscopic 
models (finite range droplet macroscopic model and folded Yukawa single 
particle microscopic model). The agreement between all the theoretical 
calculations is very remarkable and in general they are within the range 
of experimental values determined from Refs. \cite{raghavan} and \cite{raman}.
The experimental $\beta-$values from Refs. \cite{raghavan} and \cite{raman}
are represented in Fig.1 by the endpoints of the black boxes.

There is still another important experimental information relevant to
double beta decay that we wish to explore here. This is the 
$Q_{\beta\beta}$ energy of the decay. The energy released in a double 
beta process in a transition from ground-state to ground-state is given by 
\begin{equation}
Q_{\beta\beta} = \left[ M_{\rm parent} - M_{\rm daughter} -2 m_e\right] =
\left[ 2(m_n-m_p-m_e) + BE(Z,N) - BE(Z+2,N-2) \right] \, ,
\end{equation} 
in terms of the nuclear masses $M$'s, or similarly, in terms of the binding 
energies $BE$'s of parent $(Z,N)$ and daughter $(Z+2,N-2)$ nuclei.

We can see in Table 3 the experimental values of $Q_{\beta\beta}$. They are 
compared with the values calculated with the force Sk3. The agreement with 
experiment is reasonable taking into account that we are dealing with 
differences of energies ranging from 400 MeV in A=48 systems to 1200 MeV 
in A=150 and a tiny percent error in the theoretical masses may induce a 
relatively large error in $Q_{\beta\beta}$ values. From Table 3 we can see 
that to match the experimental energies, we need to increase slightly the 
energy difference in A=48,76,82,116,128,130,136 and to decrease it in 
A=96,100,150.

In this work we have used Sk3 as a representative of the Skyrme forces 
without any attempt to optimize the agreement with the experimental 
$Q_{\beta\beta}$ values. Even if Sk3 force may not be the best Skyrme 
force to accurately predict the right $Q_{\beta\beta}$ values, it is very 
interesting to see that experimental and theoretical values are in many 
cases quite close. For the cases examined here it appears that the 
$Q_{\beta\beta}$ values are fairly well reproduced for A=82,100,150 and 
that deviations from experiment in the worst case is 2.25 MeV in A=96, 
where $Q_{\beta\beta({\rm exp})}=3.35$ MeV and 
$Q_{\beta\beta({\rm Sk3})}=5.59$ MeV. In the future it will be 
interesting to test other Skyrme forces which may give better fits to 
the $Q_{\beta\beta}$ values. A case of particular concern for double beta 
decay calculations is that of A=128 for which the Sk3 force gives 
$Q_{\beta\beta}=-0.10$ MeV. Clearly, in this case further work is 
needed to make selfconsistent calculations that give the right 
$Q_{\beta\beta}$ and GT strengths. 

We also notice that, as illustrated in Fig. 1, by changing slightly the 
deformation value in the vicinity of the equilibrium deformation, one can 
change the binding energy correspondingly and get into agreement with the 
experimental $Q_{\beta\beta}$ value. This procedure could be justified 
from the point of view that, in principle, in the HF method one could 
consider several collective degrees of freedom and that the absolute 
minimum in the multidimensional landscape could correspond to a slightly 
different $\beta$ value. We can see in Fig. 1 the experimental value of 
$\delta = BE_i - BE_f$ as the distance between the two horizontal lines 
plotted in each panel. The solid horizontal line refers to the energy of 
the parent while the dashed horizontal one refers to the daughter binding 
energy. One of them is always a reference and signals the energy to keep 
fixed (parent for A=96,100,150 and daughter in A=48,76,82,116,128,130,136). 
The other line indicates the binding energy needed to reproduce the 
experimental $\delta = BE_i - BE_f$. Therefore, the cuts of this horizontal 
line with the corresponding energy curve indicates the deformations where 
this condition is satisfied. 

Fig. 2 shows the results obtained with HF-Sk3 for the GT strength 
distributions in $^{128}$Te and $^{136}$Xe, which are among the cases where 
the calculated $Q_{\beta\beta}$ is worse and the change in $\beta$ needed 
to fit $Q_{\beta\beta}$ is larger. We show the results obtained with the 
equilibrium deformations $\beta=-0.088$ in $^{128}$Te and $\beta=0.001$ in 
$^{136}$Xe as well as with the deformations that fit the $Q_{\beta\beta}$ 
values, $\beta=-0.005$ in $^{128}$Te and $\beta=0.102$ in $^{136}$Xe. 
We can see that the strength distributions obtained with both deformations 
are similar except for a small displacement in energies. In all the other 
cases the effect is even smaller and the strength distributions obtained 
with the equilibrium deformation or with the slightly changed deformation 
are practically unchanged. In the next section we show GT strengths 
obtained at the HF minimum.

In the case of the Woods-Saxon potential, where the deformation is an 
input parameter, we take the values from both Refs. \cite{raghavan} and 
\cite{raman}. Since for each nucleus, the two references give two 
different values of the $\beta-$parameter, we show in the next section 
the GT distributions obtained with the two values to take into account 
this uncertainty.

\section{Gamow-Teller strength distributions}

In this Section we show and discuss the Gamow-Teller strength distributions
obtained from different choices of the deformed mean fields and residual
interactions.

We notice that the relevant strength distributions for the double beta 
decay are the $B(GT^-)$ distribution of the parent nuclei and the 
$B(GT^+)$ distribution of daughter nuclei. 

As a general rule, the following figures showing the GT strength 
distributions are plotted versus the excitation energy of daughter nucleus. 
The distributions of the GT strength have been folded with Gaussian functions
of 1 MeV width to facilitate the comparison among the various calculations, 
so that the original discrete spectrum is transformed into a continuous 
profile. These distributions are given in units of $g_A^2/4\pi$ and one 
should keep in mind that a quenching of the $g_A$ factor, typically 
$g_{A,{\rm eff}}=(0.7-0.8)\ g_{A,{\rm free}}$, which appears squared in the 
GT strength, is expected on the basis of the observed quenching in charge 
exchange reactions.

In the case of the $B(GT^+)$ distributions, we first observe the different 
scale, which is about one order of magnitude smaller than the $B(GT^-)$ 
scale. This is a consequence of the Pauli blocking. In the nuclei considered 
here the number of neutrons $N$ is much larger than the number of protons 
$Z$. The difference between total $B(GT^-)$ and $B(GT^+)$ strengths (Ikeda 
sum rule, which is fulfilled in our calculations), is then a large number 
given by $3(N-Z)$ and practically determined by the magnitude of the 
$B(GT^-)$ strength.

We start in Fig. 3 with a discussion of the dependence of the GT strength
distributions on the coupling strength of the particle-hole residual
interaction $\chi ^{ph}_{GT}$ for a fixed value of the particle-particle 
coupling constant  $\kappa ^{pp}_{GT}=0$. The results correspond to HF with
the force Sk3 in the A=150 case. We can see on the left panel the $B(GT^-)$ 
strength distribution of the parent nucleus $^{150}$Nd and on the right 
panel the $B(GT^+)$ strength distribution of the daughter nucleus $^{150}$Sm. 
The pairing gap parameters are given in Table 1 and the deformations are 
given in Table 2. We can see in Fig. 3 how the most important effect of 
$\chi ^{ph}_{GT}$ on the $B(GT^-)$ strength distribution is a
shift of the strength toward higher excitation energies. This displacement
of the GT strength is accompanied by a reduction of the strength. This 
reduction can be more clearly appreciated on the $B(GT^+)$ strength 
distribution because the scale in this case is about two orders of magnitude
smaller, as it should be to fulfill the Ikeda sum rule
$\sum [B(GT^-)- B(GT^+)]=3(N-Z)= 90,78$ for Nd and Sm, respectively.
Therefore, the coupling constant $\chi ^{ph}_{GT}$ plays an important role
to reproduce the position of the $GT^-$ resonance.
On the other hand, the sensitivity of the GT strength distribution on the
particle-particle coupling constant  $\kappa ^{pp}_{GT}$ is not so
important as it can be seen from Fig. 4, where we can see the GT strength 
distributions for a fixed value of $\chi ^{ph}_{GT}=0.156$ MeV \cite{homma}
and for several values of  $\kappa ^{pp}_{GT}$ on the example of HF with
the force Sk3 in the A=150 case. As we can see, the position of the 
resonance does not change appreciably. Therefore, other methods, such as 
fitting the half-lives of unstable nuclei in the same mass region, have 
to be used to get phenomenologically their values.

In the next set of figures (Figs. 5-14) we show, for each couple of double 
beta decay partners, the results obtained for the $B(GT^-)$ strength 
distributions of the parent nuclei on the top panels and for the $B(GT^+)$ 
strength distributions of the daughter nuclei on the bottom panels. Also 
shown are the experimental data whenever they are available.
In each figure the left panels correspond to HF+BCS+QRPA calculations with
the force Sk3 and the right panels to WS+BCS+QRPA calculations.
In the case of HF we use the equilibrium deformations. We show with dashed 
lines the 2qp results for HF+BCS calculations where the residual interaction 
is not considered. This serves as a reference and can be used to see the 
necessity of the residual force to get into agreement with experiment. 
Solid lines are the results obtained with $\chi ^{ph}_{GT}=0.1$ MeV and 
$\kappa ^{pp}_{GT}=6/A$ MeV, which produce a good fit to all the measured
GT resonances of double beta emitters, as well as to the two-neutrino
double beta decay matrix elements, as we shall see in the next section. 
This  small value of the $\chi ^{ph}_{GT}$ coupling constant needed to 
reproduce the experimental GT resonances within a selfconsistent approach 
with Skyrme forces is in agreement with the same observation reported in 
Refs. \cite{gese,feni} and reflects the fact that one needs less residual 
interaction when using realistic effective density-dependent forces that 
when using phenomenological potentials to generate the single-particle 
energies and wave functions.

In the case of calculations with the WS potentials shown on the right hand
panels, we show results for the two different experimental deformations 
as obtained from Refs. \cite{raghavan} (solid lines) and \cite{raman} 
(dashed lines), which are given in Table 2. The calculations are done for 
a fixed value of the $\chi ^{ph}_{GT}$ and $\kappa ^{pp}_{GT}$ 
constants as obtained from the parametrization in Ref.\cite{homma}. 

Some general common features to all figures can be established first.
Concerning the HF calculations, the value of $\chi ^{ph}_{GT}$ given by 
the parametrization of Ref. \cite{homma} is an overestimation when dealing
with selfconsistent Skyrme HF calculations. Actually, a small value of
$\chi ^{ph}_{GT}=0.1$ MeV is already able to reproduce the experimental
position of the GT resonance. This is a consequence of the structure of 
the two-body density-dependent Skyrme force that contains terms like 
spin exchange operators leading to a spin-spin interaction in the 
selfconsistent mean field, which is absent in the WS potential.
The agreement with the experimental energy of the GT resonance is in
this case very good as it can be seen in the cases A=76,82,100,116,128,130,
where this information is available. Indeed, the experimental giant GT 
resonances shown in these figures represent the centroids of broad bumps.
The resonance in $^{48}$Ca reported at 10 MeV in Ref.\cite{expca48} and 
used in the fitting procedure of Ref. \cite{homma} is also in good 
agreement with our results.

With respect to the calculations performed with the WS potential, we can 
see that larger deformations produce peaks in the GT distributions 
displaced to higher energies. This is a consequence of the larger 
separation of the single particle energies when the deformation increases. 
Thus, since the deformation derived from Ref. \cite{raman} is larger than 
that of Ref. \cite{raghavan}, solid lines appear in general on the left 
of dashed lines.

It is also remarkable the good agreement with experiment obtained in 
this case. This agreement is mainly determined by the fixed value of 
$\chi ^{ph}_{GT}$ from Ref. \cite{homma}, which is still valid when 
describing the mean field with a WS potential. One should keep in mind 
that the parametrization of Ref. \cite{homma} was obtained using a Nilsson 
potential.

In Table 4 we compare the total GT strength measured and calculated with 
both HF and WS. When a standard quenching factor of 0.6 is included in the 
theoretical results fair agreement is found between theory and experiment.

\section{Two-neutrino double beta decay matrix elements}

In this section we analyze the effects of deformation, as well as the
effect of the mean field and residual interactions on the 
$2\nu\beta\beta$-decay. First, we discuss the sensitivity to deformation of 
the nuclear structure contribution $M^{2\nu}_{\rm GT}$ to the 
$2\nu\beta\beta$ half-lives. This is done in Fig. 15, where we show the 
matrix elements $M^{2\nu}_{\rm GT}$ as a function of both parent and 
daughter deformations. The figure corresponds to the decay 
$^{96}$Zr $\rightarrow$ $^{96}$Mo calculated within a Woods-Saxon scheme 
with residual interactions from Ref. \cite{homma}. We have changed freely 
the deformations of both parent and daughter nuclei without any constraint 
from experiment. In this way we can study qualitatively the effect
of deformation. The experimental values for $M^{2\nu}_{\rm GT}$ shown in 
Fig. 15, as well in the next figures have been extracted from the adopted 
experimental half-lives $T^{2\nu}_{1/2}$ given in Ref. \cite{barabash}.
From the experimental half-lives and the corresponding kinematical
factors $G^{2\nu}$, we extract two experimental nuclear matrix elements 
from Eq. (\ref{t2nu}) by assuming values for the axial coupling constant 
$g_A=1.25$ or $g_A=1$. These two values are plotted in Figs. 15-18 as 
horizontal lines.

From Fig. 15 we can see that the matrix elements $M^{2\nu}_{\rm GT}$ have 
maximum values for equal deformations of both parent and daughter and these 
values decrease rapidly when the difference between the deformations of 
parent and daughter increases.
In particular, we observe that the $M^{2\nu}_{\rm GT}$ value obtained within
a spherical picture ($\beta_{\rm parent}=\beta_{\rm daughter}=0$) is about
the upper limit and only comparable with values obtained with same 
deformations for parent and daughter in the deformed picture.
As soon as the deformations of parent and daughter change, we get a reduction
in the $M^{2\nu}_{\rm GT}$ matrix elements that cannot be obtained from
a spherical description.
The mechanism of this reduction due to the different deformations was studied 
in Ref. \cite{doublege}, where it was found that the overlap factor 
in Eqs.(\ref{m2nu}-\ref{bcsif}) is at the origin of the suppression.
We can see in Fig. 15 that the experimental values of $M^{2\nu}_{\rm GT}$,
plotted as thick segments in each curve, are compatible with particular 
values of parent and daughter deformations.

We show in Fig. 16 the difference between parent and daughter nuclear
quadrupole deformations for the double-beta emitters.
The dots correspond to the results obtained from selfconsistent HF
calculations with the Skyrme force Sk3, while the extreme values on
the vertical segments indicate the maximum and minimum differences
compatible with the experimental values in Table 2. These are also
the extreme values used in WS calculations in Fig. 17.

In the next two figures (Figs. 17-18) we show the matrix elements 
$M^{2\nu}_{\rm GT}$ for the same double beta emitters studied in the last 
section as a function of the particle-particle strength $\kappa ^{pp}_{GT}$.
The experimental values extracted by assuming $g_A=1.25$ or $g_A=1$ are
shown by the lower and upper horizontal lines, respectively.

It is well known \cite{2bqrpa} that the $pp$ interaction introduces
a different mechanism of suppression of the $M^{2\nu}_{\rm GT}$ 
matrix elements, which is also interesting to study in our case.
In this way we can compare the effect of the $pp$ force with the effect
due to deformation.
We first discuss in Fig. 17 the effects of deformation in the WS case
by taking the available experimental quadrupole deformations and then
we discuss the HF case by considering the selfconsistent deformations.

In the WS case (Fig. 17) we use the same potential parameters, gaps and $ph$ 
residual interaction as those used in the single beta calculations in the 
previous section. 
For deformations we take all the experimental possibilities for parent and 
daughter given in Table 2 and cross them to calculate $M^{2\nu}_{\rm GT}$. 
Then we show in the figure the upper and lower results obtained as a function 
of $\kappa ^{pp}_{GT}$ and we draw a shadow region between them.
We also show for comparison the results obtained in the spherical case (dashed
lines).

The first thing to notice is the already mentioned reduction of 
$M^{2\nu}_{\rm GT}$ as the magnitude of $\kappa ^{pp}_{GT}$ increases, which 
takes place for both spherical and deformed cases, although the effect of 
the $pp$ force is larger in the spherical case. The spherical curves decay 
faster than the deformed ones with $\kappa ^{pp}_{GT}$. This means that the 
deformed results are more stable (more insensitive) to the particular 
strength of the $pp$ interaction.

Another interesting feature to mention is that, as expected from the analysis 
in Fig. 15, deformation introduces in most cases a reduction factor with 
respect to the spherical result. Only when the deformations of parent and 
daughter are very similar, the results obtained in the deformed case can be
larger than the spherical ones. This is for instance the case of $A=82$
(see Table 2), where the experimental quadrupole deformations \cite{raman} 
are $\beta=0.1944$ in the case of the parent nucleus $^{82}$Se and  
$\beta=0.2022$ in the case of the daughter nucleus $^{82}$Kr.

In Fig. 18 we show the results corresponding to HF calculations with the
force Sk3. In this case the quadrupole deformations are obtained 
selfconsistently (see Table 2) and are the same we used in the previous
section to calculate the GT strength distributions in the single 
$\beta-$decays. The coupling strength of the $ph$ residual interaction has 
been taken $\chi=0.1$ MeV as in the previous section. Contrary to the case of 
single $\beta-$decay, where the position of the GT resonance is determined
by the strength of the $ph$ force and almost independent on the 
$\kappa ^{pp}_{GT}$ force, we can see in Fig. 18 the sensitivity of the 
$2\nu\beta\beta$ decay to the $pp$ force.
We find that the experimental values of $M^{2\nu}_{\rm GT}$ are roughly
reproduced with values of $\kappa ^{pp}_{GT}=6/A$ MeV. This is the reason
why we used these values also in the HF calculations of the GT distributions 
in the previous section.
  
To illustrate even further the effect of deformation on the $M^{2\nu}_{\rm GT}$
matrix elements, we show in Fig. 19 the HF+BCS+QRPA results with the Skyrme
force SG2 for the decay 
$^{96}$Zr $\rightarrow$ $^{96}$Mo corresponding to the decay of a prolate 
parent to a prolate daughter or to the decay of a spherical parent into
a spherical or a prolate daughter. The actual deformations used in these
calculations are the consistently obtained with the force SG2 as can
be seen in Table 2, namely $\beta =0.016,\, 0.147$ for the parent
$^{96}$Zr and $\beta =-0.006,\, 0.119$ for the daughter $^{96}$Mo.
We can see that the transition from the spherical shape to the prolate
shape reduces considerably the matrix elements as compared to the
spherical/spherical or to the prolate/prolate cases, which are comparable.
This reduction due to the different deformations makes the results 
compatible with the experimental values.

\section{Concluding remarks}

Using a deformed QRPA formalism, which includes $ph$ and $pp$ separable 
residual interactions, we have studied the GT strength distributions for 
the two decay branches $\beta^-$ and $\beta^+$ in double beta decay 
processes, as well as the two-neutrino double beta decay matrix elements. 
In the same manner in which two-neutrino double beta decay is used to 
calibrate the nuclear matrix elements for neutrinoless double beta 
decay, the single beta decay branches of parent and daughter are 
used to test the matrix elements for two-neutrino double beta decay. 

Two different methods, a deformed Skyrme HF approach and a phenomenological 
deformed WS potential, are used to construct the quasiparticle mean field, 
which includes pairing correlations in BCS approximation. 

In the case of HF the deformation is determined selfconsistently and we are 
able to reproduce the experimental charge radii and quadrupole moments. In 
the case of WS the input deformation is taken from experiment and we use 
two values for each nucleus, one corresponding to the measured quadrupole 
moment of the first $2^+$ state and the other extracted from the measured 
$B(E2)$ values. The latter can be considered as an upper limit of the 
$\beta-$value. More experimental work would be needed to improve and 
complete the experimental determination of the quadrupole moments based 
on the first of these methods.

Starting from these quasiparticle basis we perform a pnQRPA calculation
with separable forces to obtain the energy distributions of the $GT^-$ 
strength in the parent nucleus and the $GT^+$ strength in the daughter
nucleus and from them the $2\nu\beta\beta-$decay matrix elements.
It is well known from previous studies that the $ph$ force allows
a reasonable fit of the GT resonance, and that the $pp$ force also affects
the $B(GT^-)$ and $B(GT^+)$ distributions. This knowledge has been applied 
in our paper to test existing parametrizations of the $ph$ and $pp$ residual 
forces and to see how good agreement can be obtained between experiment and
theory with HF and WS methods. To our knowledge, we have considered for the 
first time simultaneously all the possible two-beta emitters and their 
corresponding daughters comparing their $B(GT)$ to experiment.
We find that we need different strengths of the $ph$ force to reproduce 
the position of the GT resonance, depending on the HF or WS basis. In the 
first case, a small value of $\chi ^{ph}_{GT}=0.1$ MeV reproduces all the 
measured GT resonances. In the second case the parametrization obtained 
in Ref. \cite{homma} ($\chi ^{ph}_{GT}= 5.2 \; /A^{0.7}$ MeV), using a 
Nilsson potential, is still valid when using a WS potential. The fact 
that $\chi ^{ph}_{GT}$ is smaller in HF than in WS can be understood as 
arising from the fact that the HF mean field already contains the average 
effect of spin-spin interaction terms. In both cases we reproduce 
reasonably well not only the position of the resonances but also the total 
GT strength. It should also be mentioned that the $GT^-$ strength of the 
parent nucleus and the $GT^+$ strength of the daughter are located at 
different energies, a feature that is relevant for double beta decay 
because it introduces a reduction of the double beta decay probabilities.
It would be very useful to improve and complete the experimental 
information on GT strength distributions by (p,n) and (n,p) charge 
exchange reactions on nuclei participating in double beta decay.

We have also explored the theoretical $Q_{\beta\beta}$ values obtained 
with the HF method. We find that with the Sk3 force used here the agreement 
with the experimental $Q_{\beta\beta}$ is not perfect and that it will be 
worth to look for a Skyrme force that may optimize agreement with experiment 
on both GT strengths and $Q_{\beta\beta}$ values. 
However, taking into account that there is no fitting parameter at
all, the agreement between theory and experiment is good. So far, no 
other approach used to study double beta decay has been able to obtain 
$Q_{\beta\beta}$ values so close to experiment. This is an important 
advantage of the HF method that deserves further exploration. 

The effect of deformation on the  $2\nu\beta\beta-$decay matrix elements
has been studied first by considering the deformations of both parent and 
daughter as free parameters. It is found that the matrix elements are
suppressed with respect to the spherical case. More precisely, it is found
a sizable reduction effect that scales with the deformation difference 
between parent and daughter. This suppression mechanism, which is ignored
in spherical treatments, may play an important role in approaching the
theoretical estimates to experiment.

In the case of the WS potential, we have studied the $2\nu\beta\beta-$decay 
matrix elements by considering the maximum and minimum differences between
the experimental deformations of parent and daughter.
This procedure generates a region of decaying rates
that would be narrowed from an improved experimental determination of the 
quadrupole deformations.

In the case of HF calculations we find that using the selfconsistent
deformations obtained from the minimization of the energy and 
residual interactions with coupling strengths given by
$\chi ^{ph}_{GT}=0.1$ MeV and  $\kappa ^{pp}_{GT}=6/A$ MeV, we
are able to reproduce simultaneously the available experimental
information on the GT strength distributions of the single $\beta-$branches
and the  $2\nu\beta\beta-$decay matrix elements.

\begin{center}
{\Large \bf Acknowledgments} 
\end{center}
We would like to thank A.A. Raduta for useful discussions. This work was 
supported by Ministerio de Educaci\'on y Ciencia (Spain) under contract 
number BFM2002-03562 and by International Graduiertenkolleg GRK683, by 
the ``Land Baden-Wuerttemberg'' within the ``Landesforschungsschwerpunkt: 
Low Energy Neutrinos'', by the DFG under 418SPA112/8/03, the VEGA Grant 
agency of the Slovac Republic under contract No.~1/0249/03, and by the 
EU ILIAS project under contract RII3-CT-2004-506222. One of us (R.A.-R.) 
thanks Ministerio de Educaci\'on y Ciencia (Spain) for financial support. 

\vfill\eject

\newpage

\begin{center}

\begin{table}[t]
\caption{Pairing gaps for protons and neutrons $\Delta_p,\, \Delta_n$ 
(MeV) and charge radii $r_c$ (fm).}
\vskip 0.5cm
\begin{tabular}{ccccccc}
Nucleus & $\Delta_p$ & $\Delta_n$ & exp $r_c$ \cite{radiiexp} & $r_c$ Sk3 
& $r_c$ SG2 & $r_c$ \cite{ring} \\ \\
\hline \\
$^{48}$Ca & 2.18 & 1.68 & 3.4736(8)  & 3.586 & 3.549 & 3.471 \\
$^{48}$Ti & 1.90 & 1.56 &  3.592     & 3.628 & 3.597 & 3.571 \\ \\
$^{76}$Ge & 1.56 & 1.54 & 4.127(8)   & 4.130 & 4.083 & 4.057 \\
$^{76}$Se & 1.75 & 1.71 & 4.152(9)   & 4.170-4.180 & 4.113-4.143 & 4.119 \\ \\
$^{82}$Se & 1.41 & 1.54 & 4.122(8)   & 4.204 & 4.159 & 4.131 \\
$^{82}$Kr & 1.72 & 1.64 & 4.1921(11) & 4.196 & 4.196 & 4.173 \\ \\
$^{96}$Zr & 1.53 & 0.84 & 4.3508(12) & 4.433-4.443 & 4.342-4.389 & 4.376 \\
$^{96}$Mo & 1.53 & 1.03 & 4.377(10)  & 4.448-4.457 & 4.369-4.388 & 4.381 \\ \\
$^{100}$Mo &1.60 & 1.36 & 4.447(10)  & 4.516 & 4.439-4.466 & 4.448 \\
$^{100}$Ru &1.55 & 1.30 & 4.453 & 4.516 & 4.457 & 4.449 \\ \\
$^{116}$Cd &1.47 & 1.37 & 4.625 & 4.703-4.715 & 4.653 & 4.643 \\
$^{116}$Sn &1.77 & 1.20 & 4.625 & 4.709-4.753 & 4.702 & 4.609 \\ \\
$^{128}$Te &1.13 & 1.28 & 4.735 & 4.803-4.805 & 4.746 & 4.732 \\
$^{128}$Xe &1.32 & 1.27 & 4.776 & 4.836-4.839 & 4.782-4.786 &4.778 \\ \\
$^{130}$Te &1.06 & 1.18 & 4.742 & 4.812-4.816 & 4.750 & 4.739 \\
$^{130}$Xe &1.31 & 1.25 & 4.783 & 4.845-4.846 & 4.796-4.801 & 4.784 \\ \\
$^{136}$Xe &0.98 & 1.44 & 4.799 & 4.878 & 4.815 & 4.804 \\
$^{136}$Ba &1.27 & 1.03 & 4.833 & 4.902 & 4.847 & 4.837 \\ \\
$^{150}$Nd &1.23 & 1.05 & 5.047 & 5.114 & 5.055 & 5.046 \\
$^{150}$Sm &1.44 & 1.19 & 5.047 & 5.108 & 5.046 & 5.047 \\
\end{tabular}
\end{table}

\begin{table}[t]
\caption{ Theoretical and experimental $\beta$ values, see text.}
\label{table.1}
\begin{tabular}{cccccccc}\\
   & \multicolumn{2}{c}{Exp} &&  \multicolumn{4}{c}{Theory}  \\ 
\cline {2-3} 
\cline{5-8} \\
 Nucleus  & Ref. \cite{raghavan} & Ref. \cite{raman} && this work (Sk3) & 
this work (SG2)  & Ref. \cite{ring}  & Ref. \cite{moeller} \\ \\
\hline
\\
$^{48}$Ca  & 0.000    & 0.101(17) && -0.002 & -0.001 & 0.000  & 0.000 \\
$^{48}$Ti  &+0.17(10) & 0.269(7)  && -0.002 & -0.003 & -0.009 & 0.000 \\  \\
$^{76}$Ge  &+0.095(30)& 0.2623(39)&&  0.161 &  0.157 &  0.157 & 0.143  \\ 
$^{76}$Se  &+0.163(33)& 0.3090(37)&& -0.181 [+0.157] & -0.191 [+0.049]  
                &-0.244 & -0.241  \\ \\
$^{82}$Se  &+0.104(32)& 0.1944(26)&& 0.126 & 0.150 & 0.133 & 0.154 \\
$^{82}$Kr  &          & 0.2022(45)&& 0.106 & 0.103 & 0.119 & 0.071 \\ \\
$^{96}$Zr  &          & 0.081(16)&& 0.207 [-0.167] & 0.016 [+0.147] 
           &  0.223 & 0.217 \\  
$^{96}$Mo  &  +0.068(27) & 0.1720(16) & & 0.147 [-0.164] & -0.006 [+0.119]
           &   0.167  &   0.080  \\ \\
$^{100}$Mo &  +0.139(30) &  0.2309(22)&& 0.236 & 0.167 [-0.191] 
           &  0.253 & 0.244  \\
$^{100}$Ru &  +0.136(22) &  0.2172(22)&& 0.175 & 0.157 &  0.194 &  0.161  \\ \\  
$^{116}$Cd &  +0.113(11) &  0.1907(34)&& 0.206 [-0.207] & 0.209 
           &  -0.258 &  -0.241 \\   
$^{116}$Sn &  +0.043(10) &  0.1118(16)&& 0.264 [-0.134] & 0.251 [-0.034] 
           &   0.003     &  0.000  \\ \\  
$^{128}$Te &  +0.011(10) &  0.1363(11)  && -0.088 [+0.102] & 0.094 [-0.091]
           &  -0.002  &    0.000    \\ 
$^{128}$Xe &          &  0.1837(49)  && 0.148 [-0.122] & 0.150 [-0.133]
           &  0.160  &    0.143  \\ \\   
$^{130}$Te & +0.035(23) &  0.1184(14) && -0.076 [+0.051] & -0.039 [+0.066]
           & 0.032   &    0.000 \\  
$^{130}$Xe &         & 0.169(6) && 0.108 [-0.098] & 0.161 [-0.132]
           &   0.128  &   -0.113 \\ \\  
$^{136}$Xe &          & 0.086(19)&&  0.001 & 0.016 
           & -0.001  &    0.000 \\  
$^{136}$Ba &         & 0.1242(8) && 0.009 & 0.070 &  -0.002  & 0.000 \\ \\ 
$^{150}$Nd & +0.367(86) & 0.2848(21)  && 0.266 & 0.271 &  0.221 & 0.243 \\  
$^{150}$Sm & +0.230(30) & 0.1931(22)  && 0.207 & 0.203 &  0.176 & 0.206 \\   
\end{tabular}
\end{table}

\newpage

\begin{table}[t]
\caption{ Experimental and theoretical $Q_{\beta\beta}$ (MeV) values 
obtained with the Skyrme force Sk3.}
\begin{tabular}{crr} \\
 Double beta transition  & ($Q_{\beta\beta})_{\rm exp}$ &
 ($Q_{\beta\beta})_{\rm Sk3}$  \\
\hline
\\
$^{48}$Ca $\rightarrow$ $^{48}$Ti   & 4.272 &  2.95 \\
$^{76}$Ge $\rightarrow$ $^{76}$Se   & 2.039 &  1.36 \\
$^{82}$Se $\rightarrow$ $^{82}$Kr   & 2.995 &  2.58 \\
$^{96}$Zr $\rightarrow$ $^{96}$Mo   & 3.350 &  5.59 \\
$^{100}$Mo $\rightarrow$ $^{100}$Ru  & 3.034 &  3.57 \\
$^{116}$Cd $\rightarrow$ $^{116}$Sn  & 2.805 &  1.88 \\
$^{128}$Te $\rightarrow$ $^{128}$Xe  & 0.867 & -0.10 \\
$^{130}$Te $\rightarrow$ $^{130}$Xe  & 2.529 &  1.20 \\
$^{136}$Xe $\rightarrow$ $^{136}$Ba  & 2.468 &  0.80 \\
$^{150}$Nd $\rightarrow$ $^{150}$Sm  & 3.367 &  3.59 \\
\end{tabular}
\end{table}

\vskip 1cm

\begin{table}[t]
\caption{ Experimental and calculated (HF-Sk3 and WS) summed GT strength.
A standard quenching factor 0.6 has been included in the theoretical results.}
\begin{tabular}{ccccc} \\
  && exp & HF-Sk3 & WS \\
\hline \\
$\sum B(GT^+)$ & $^{48}$Ti   & $1.42 \pm 0.2$ & 1.00 & 1.79 \\
& $^{76}$Se   & $1.45 \pm 0.07$ & 0.48 & 2.06  \\ \\
$\sum B(GT^-)$ & $^{76}$Ge   & 19.89 & 21.78 & 22.65 \\
&  $^{82}$Se   & 21.91 & 25.34 & 26.09 \\
&  $^{100}$Mo  & 26.69 & 29.14 & 29.93 \\
&  $^{116}$Cd  & 32.70 & 34.79 & 36.41 \\
&  $^{128}$Te  & 40.08 & 43.22 & 43.44 \\
&  $^{130}$Te  & 45.90 & 46.85 & 46.66 \\
\end{tabular}
\end{table}

\end{center}

\newpage

\begin{center}

\begin{figure}[t]
\psfig{file=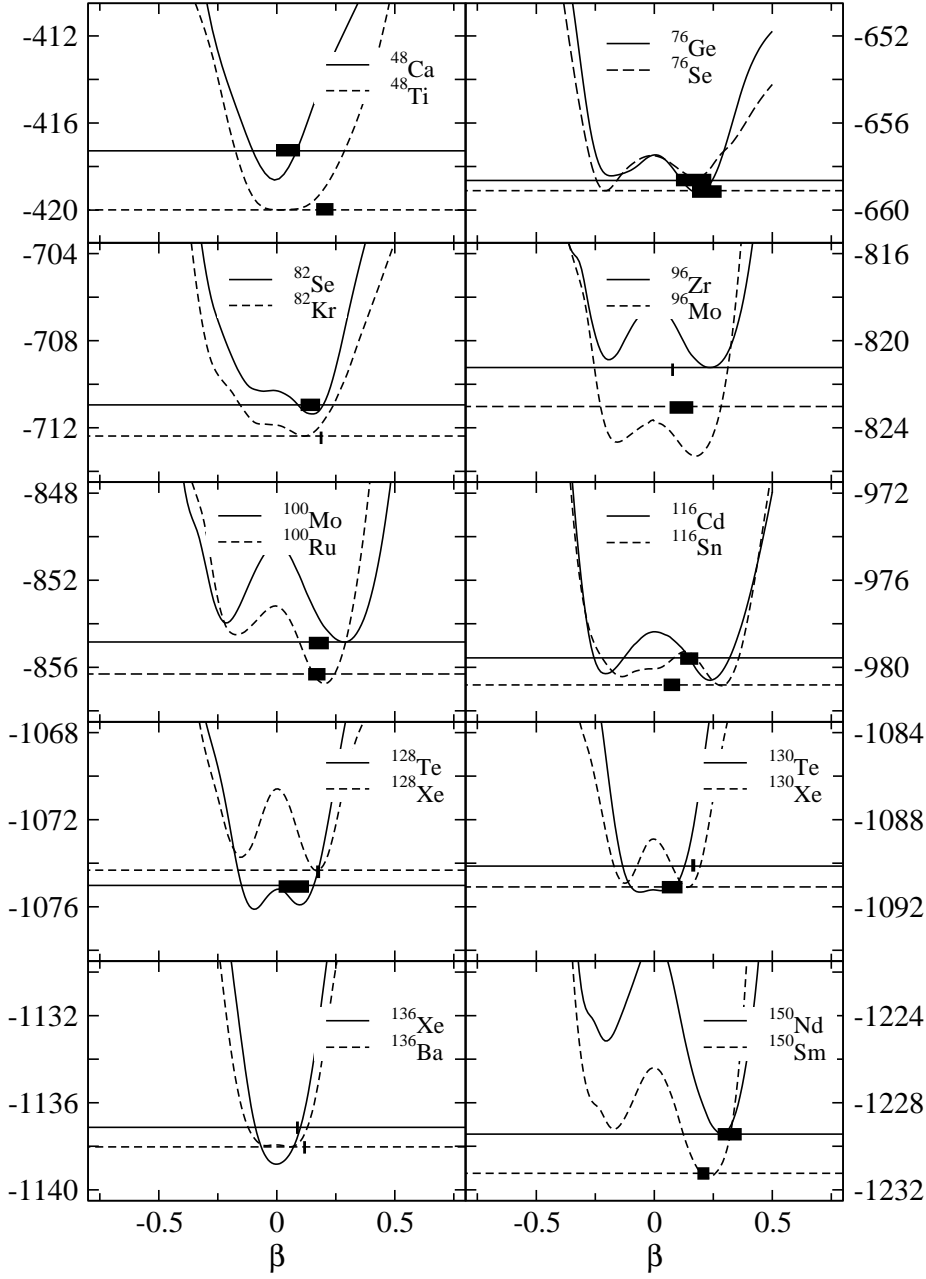,width=0.8\textwidth}
\caption{Binding energy [MeV] as a function of the quadrupole deformation 
parameter $\beta$ obtained from deformed Hartree-Fock calculations with the 
Skyrme force Sk3. Experimental $\beta$ values from Refs. 
\protect\cite{raghavan} and \protect\cite{raman} are represented as the 
extreme values of the black boxes. The experimental difference of binding 
parent and daughter binding energies are given by the distance between the
two horizontal lines (see text).}
\end{figure}

\newpage

\begin{figure}[t]
\psfig{file=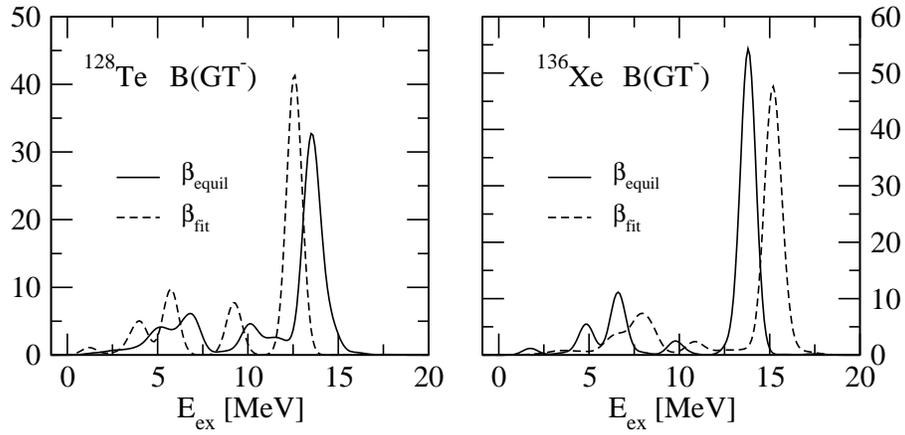,width=0.8\textwidth}
\caption{HF-Sk3 $B(GT^-)$ strength distributions $[g_A^2/4\pi]$ in $^{128}$Te 
and $^{136}$Xe calculated with the equilibrium deformation ($\beta_{\rm equil}$)
and with the deformation that fits the experimental $Q_{\beta\beta}$ values
($\beta_{\rm fit}$).}
\end{figure}

\newpage

\begin{figure}[t]
\psfig{file=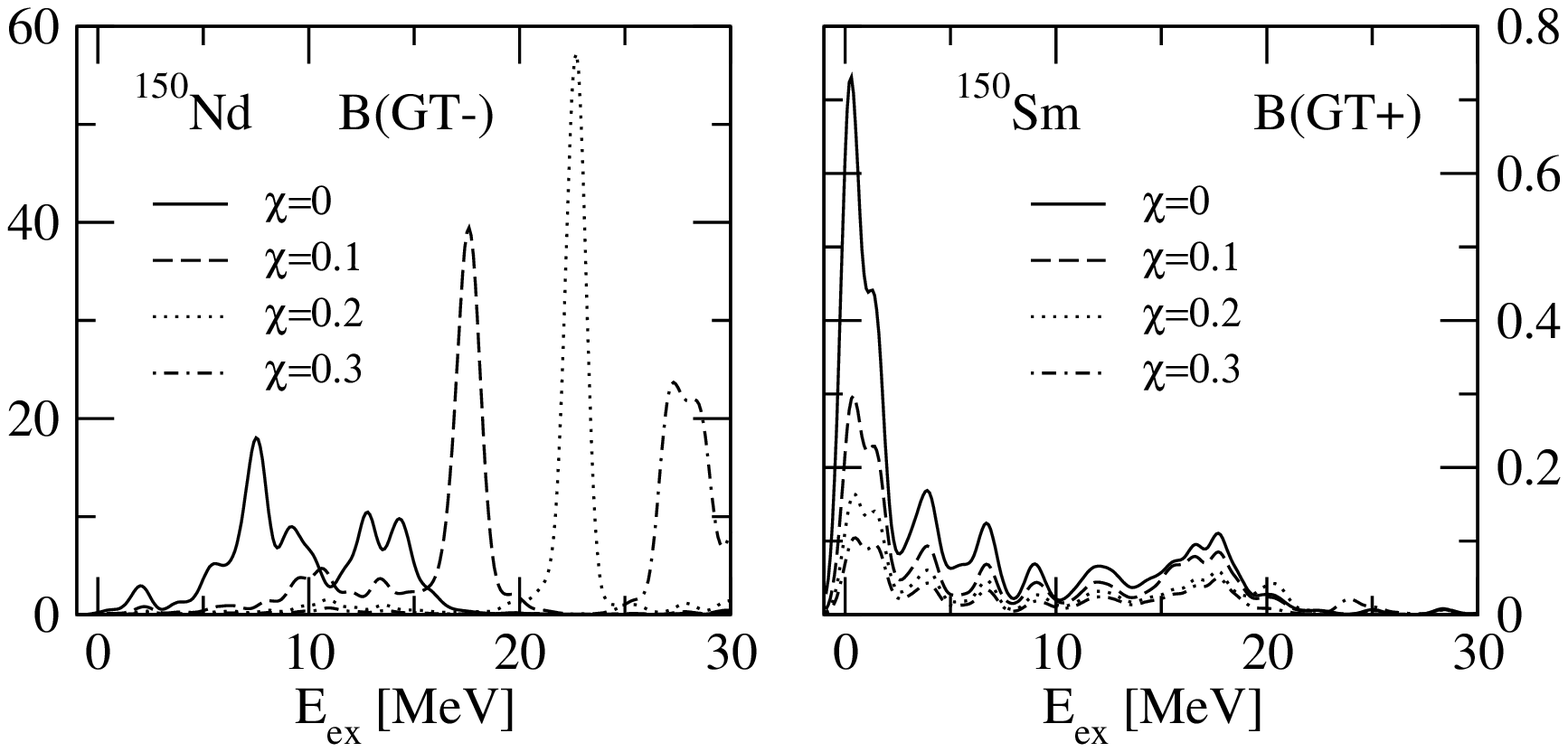,width=0.8\textwidth}
\caption{HF-Sk3 Gamow-Teller strength distributions $[g_A^2/4\pi]$ in $^{150}$Nd 
and $^{150}$Sm for various values of the coupling strength $\chi ^{ph}_{GT}$ 
[MeV].}
\end{figure}

\newpage

\begin{figure}[t]
\psfig{file=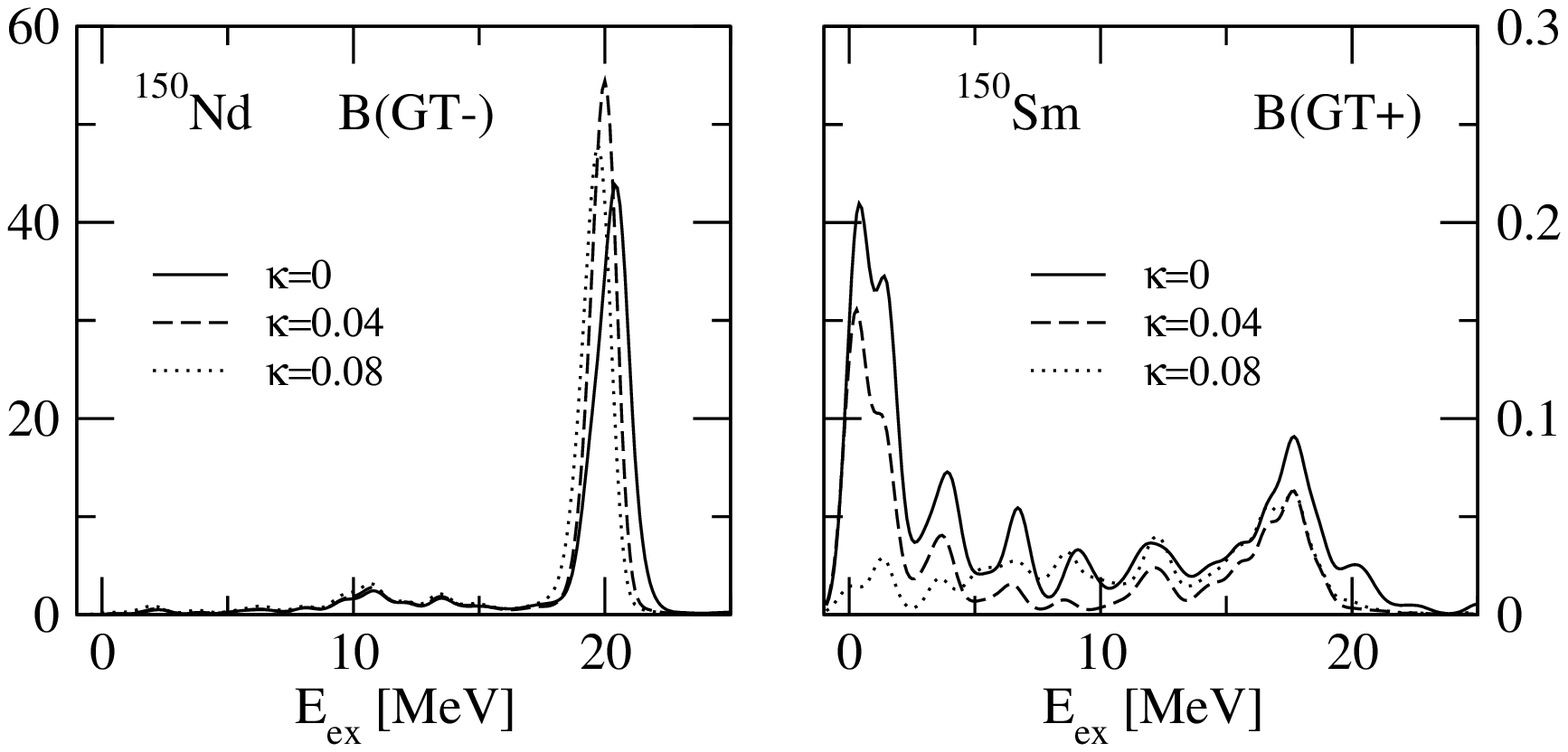,width=0.8\textwidth}
\caption{HF-Sk3 Gamow-Teller strength distributions $[g_A^2/4\pi]$ in $^{150}$Nd 
and $^{150}$Sm for various values of the coupling strength $\kappa ^{pp}_{GT}$ 
[MeV].}
\end{figure}

\newpage

\begin{figure}[t]
\psfig{file=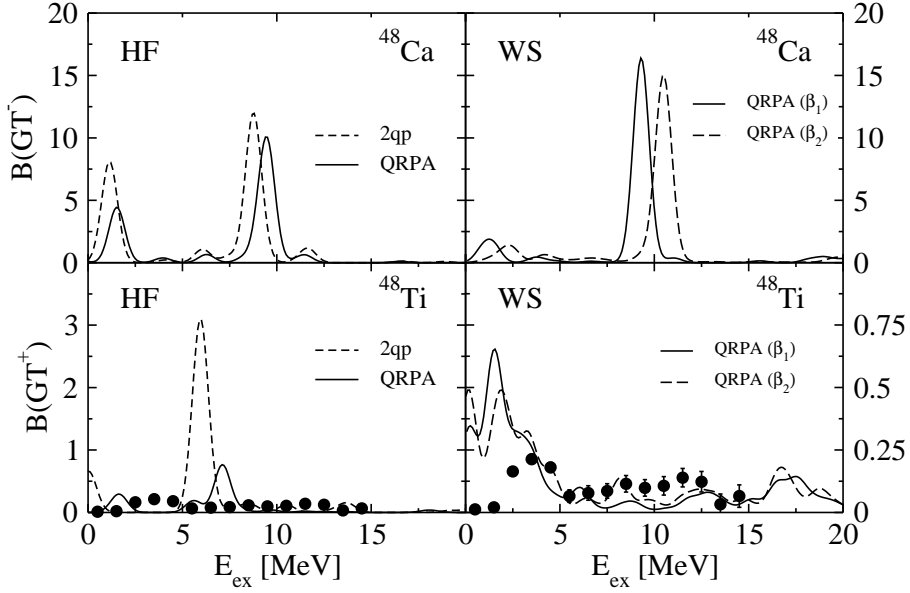,width=0.8\textwidth}
\caption{Gamow-Teller $B(GT^-)$ and $B(GT^+)$ strength distributions
$[g_A^2/4\pi]$ in $^{48}$Ca and $^{48}$Ti plotted as a function of the
excitation energies of the corresponding daughter nuclei. Left panels 
show results from HF(Sk3) calculations without residual interaction
(dashed lines) and with residual interactions with $\chi ^{ph}_{GT}=0.10$ MeV,  
$\kappa ^{pp}_{GT}=6/A$ MeV (solid lines). Right panels show results using 
WS potentials with $\chi ^{ph}_{GT}$ and $\kappa ^{pp}_{GT}$ from 
Ref.\protect\cite{homma} and with two different values for the quadrupole 
deformation: $\beta_1$ from Ref. \protect\cite{raghavan} (solid line) and  
$\beta_2$ from Ref. \protect\cite{raman} (dashed line). Experimental data are 
from Ref.\protect\cite{exp48ti}. Notice that no quenching factor has been 
included in the calculations.}
\end{figure}

\newpage

\begin{figure}[t]
\psfig{file=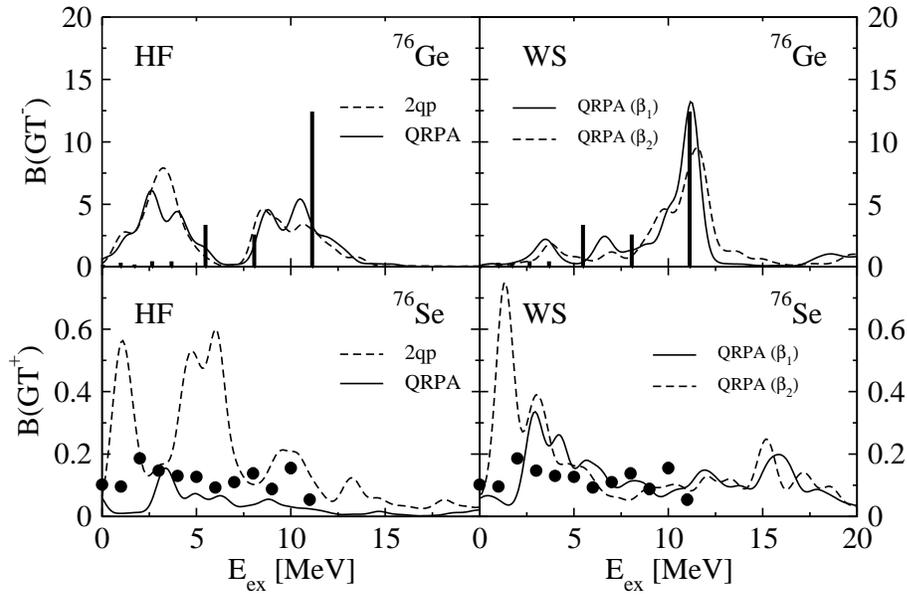,width=0.8\textwidth}
\caption{Same as in Fig. 5 for $^{76}$Ge and $^{76}$Se. Data in $^{76}$Se
are from \protect\cite{seexp}. Vertical lines in $^{76}$Ge are experimental 
data from \protect\cite{expmadey}. }
\end{figure}

\newpage

\begin{figure}[t]
\psfig{file=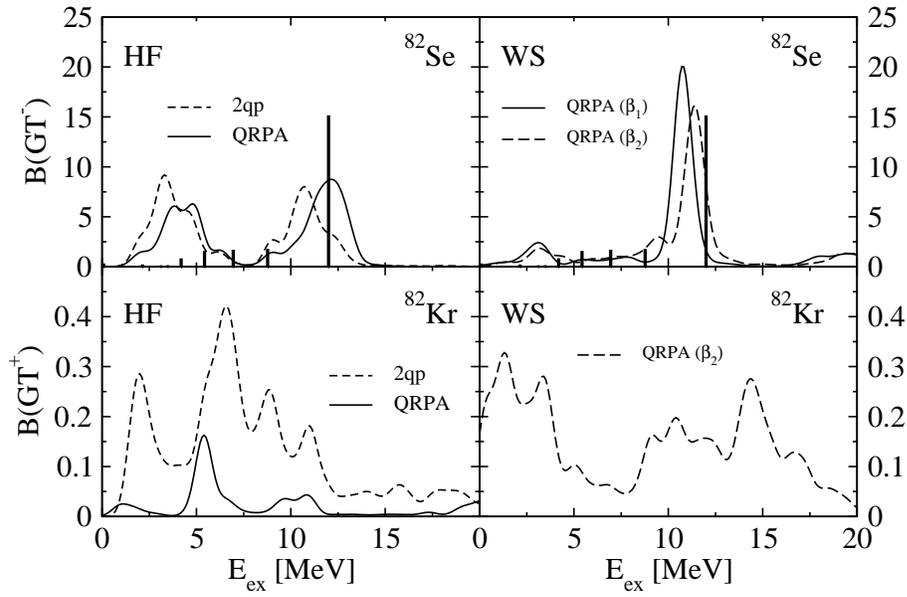,width=0.8\textwidth}
\caption{Same as in Fig. 5 for $^{82}$Se and $^{82}$Kr. Vertical lines in 
$^{82}$Se are experimental data from \protect\cite{expmadey}.}
\end{figure}

\newpage

\begin{figure}[t]
\psfig{file=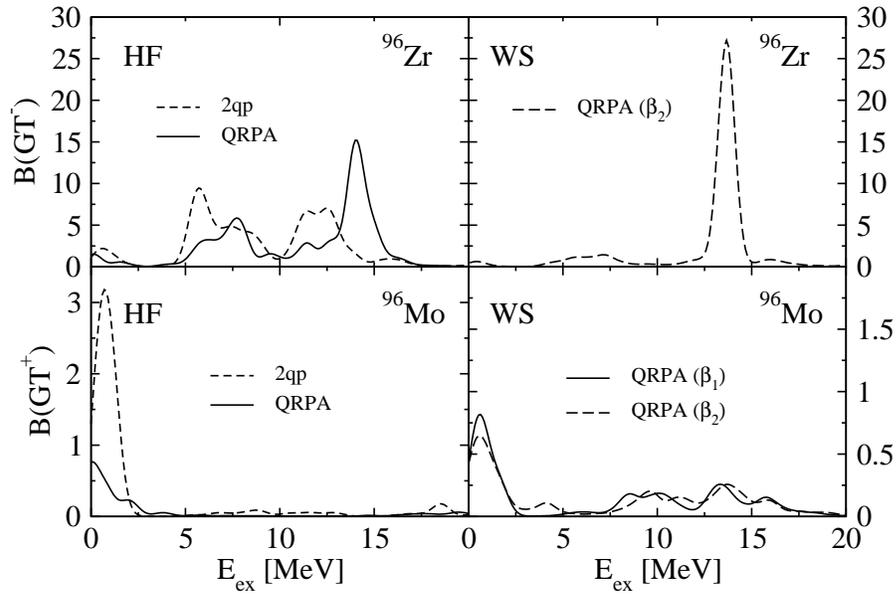,width=0.8\textwidth}
\caption{Same as in Fig. 5 for $^{96}$Zr and $^{96}$Mo.}
\end{figure}

\newpage

\begin{figure}[t]
\psfig{file=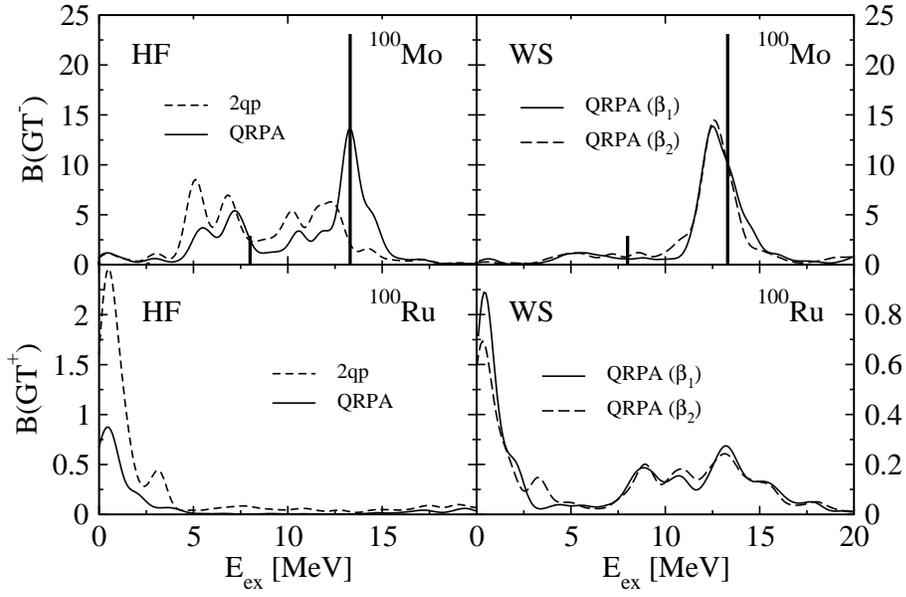,width=0.8\textwidth}
\caption{Same as in Fig. 5 for $^{100}$Mo and $^{100}$Ru. Vertical lines in 
$^{100}$Mo are experimental data from \protect\cite{expmocd}.}
\end{figure}

\newpage

\begin{figure}[t]
\psfig{file=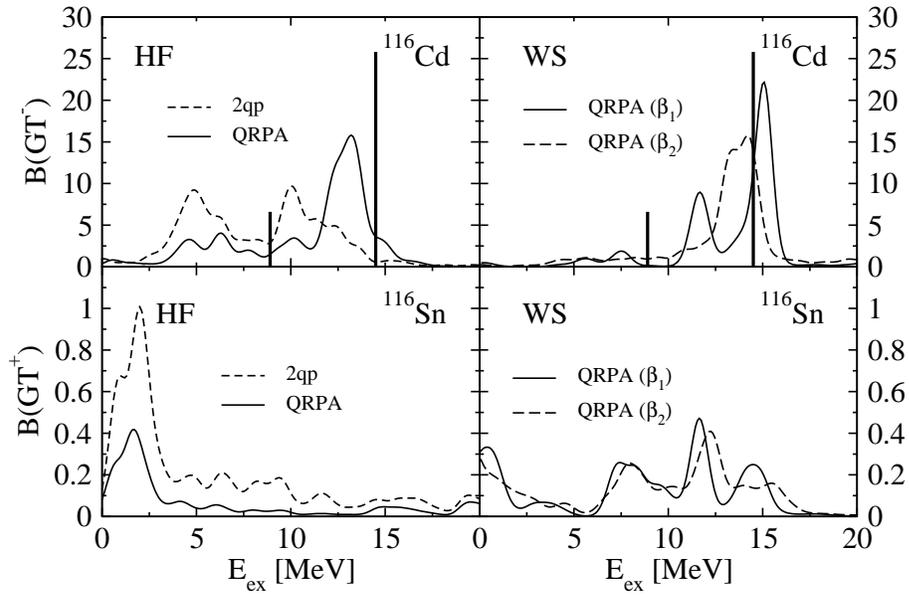,width=0.8\textwidth}
\caption{Same as in Fig. 5 for $^{116}$Cd and $^{116}$Sn. Vertical lines in 
$^{116}$Cd are experimental data from \protect\cite{expmocd}.}
\end{figure}

\newpage

\begin{figure}[t]
\psfig{file=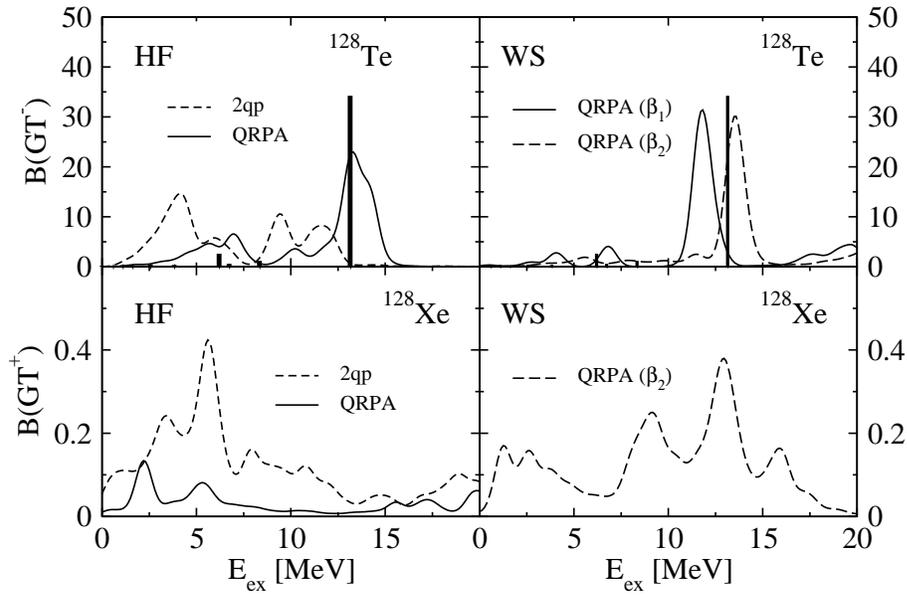,width=0.8\textwidth}
\caption{Same as in Fig. 5 for $^{128}$Te and $^{128}$Xe. Vertical lines in 
$^{128}$Te are experimental data from \protect\cite{expmadey}.}
\end{figure}

\newpage

\begin{figure}[t]
\psfig{file=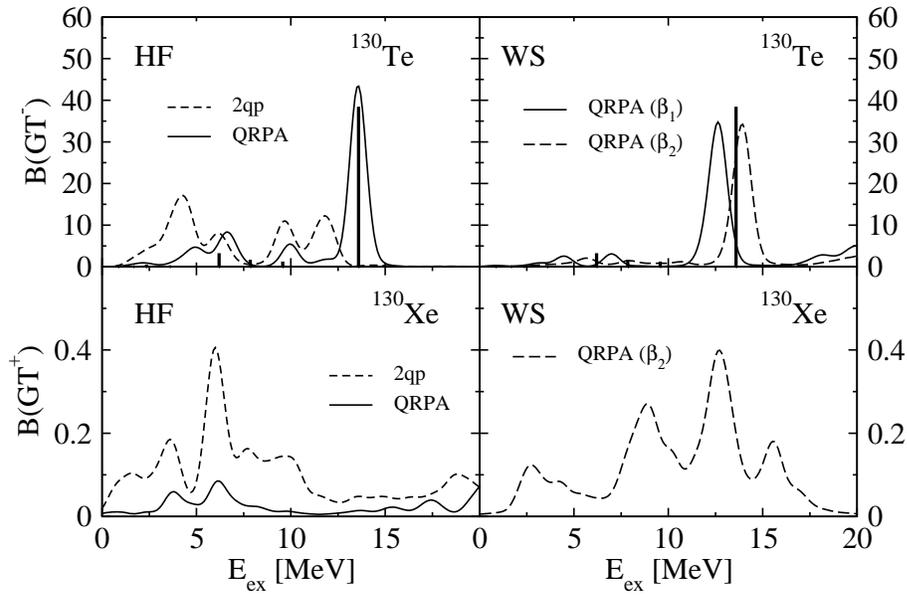,width=0.8\textwidth}
\caption{Same as in Fig. 5 for $^{130}$Te and $^{130}$Xe. Vertical lines in 
$^{130}$Te are experimental data from \protect\cite{expmadey}.}
\end{figure}

\newpage

\begin{figure}[t]
\psfig{file=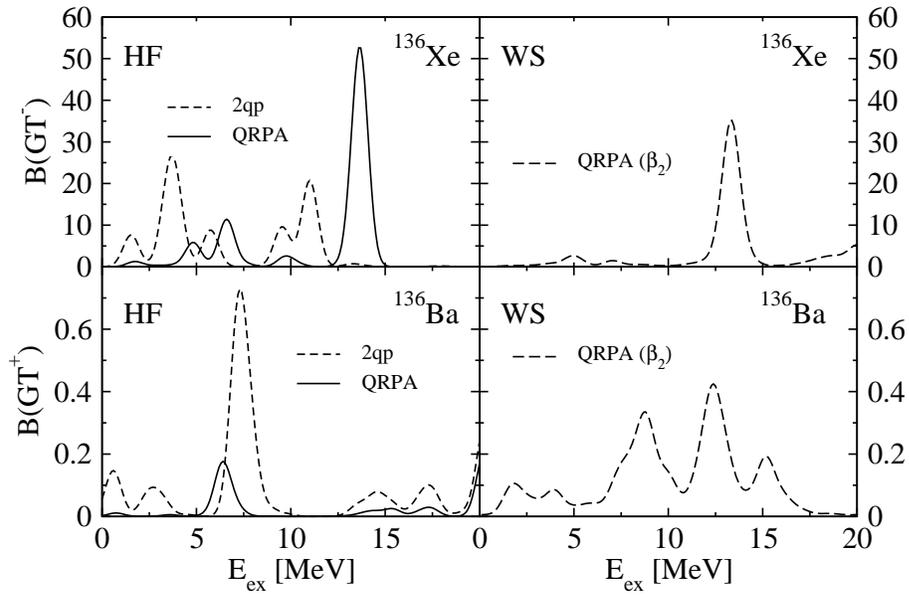,width=0.8\textwidth}
\vskip 1cm
\caption{Same as in Fig. 5 for $^{136}$Xe and $^{136}$Ba.}
\end{figure}

\newpage

\begin{figure}[t]
\psfig{file=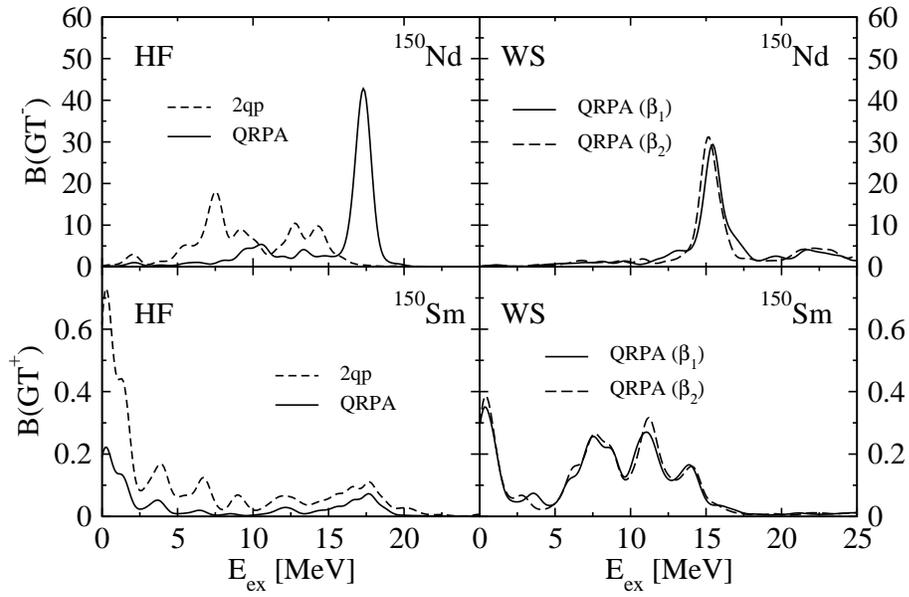,width=0.8\textwidth}
\caption{Same as in Fig. 5 for $^{150}$Nd and $^{150}$Sm.}
\end{figure}

\newpage

\begin{figure}[t]
\psfig{file=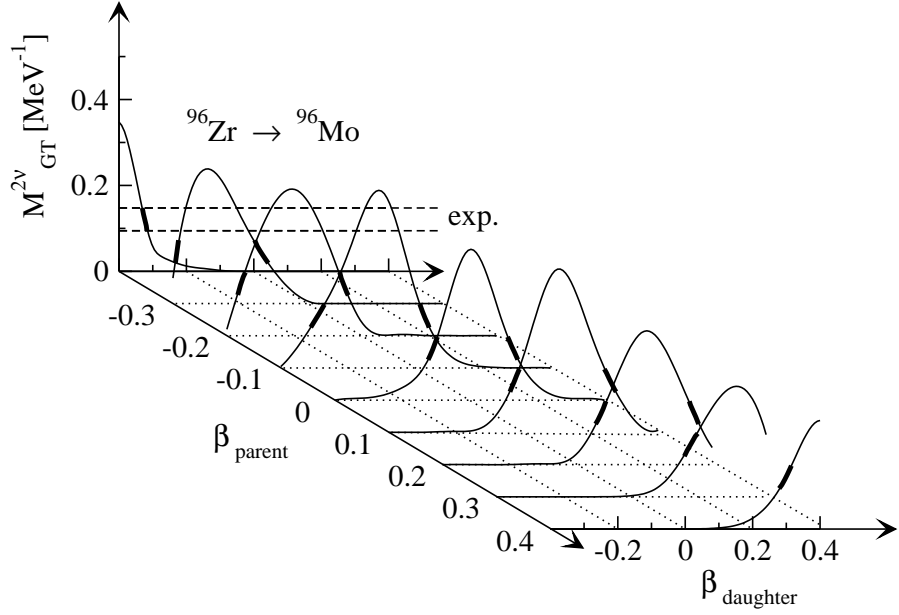,width=0.8\textwidth}
\caption{$2\nu\beta\beta-$decay matrix elements of $^{96}$Zr as a function
of both parent and daughter deformations. The two dashed horizontal lines
correspond to experimental $M^{2\nu}_{\rm GT}$ extracted from Ref. 
\protect\cite{barabash} using $g_A=1.0$ and $g_A=1.25$. The thick segments 
in each curve correspond to the experimental values of $M^{2\nu}_{\rm GT}$.}
\end{figure}

\newpage

\begin{figure}[t]
\psfig{file=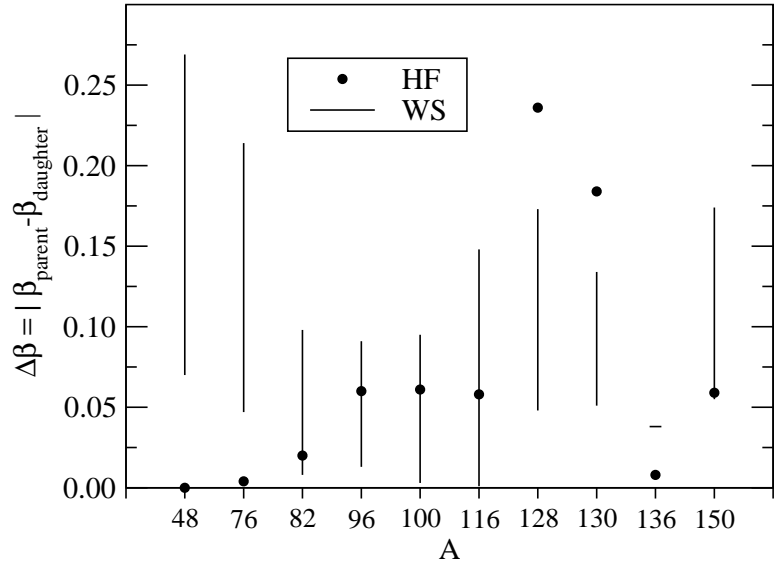,width=0.8\textwidth}
\caption{Difference between parent and daughter quadrupole deformations
in double-beta emitters. Dots are selfconsistent results from Skyrme Sk3
calculations. Vertical lines indicate the maximum and minimum experimental
differences (see Table 2), which are used in WS calculations.}
\end{figure}

\newpage

\begin{figure}[t]
\psfig{file=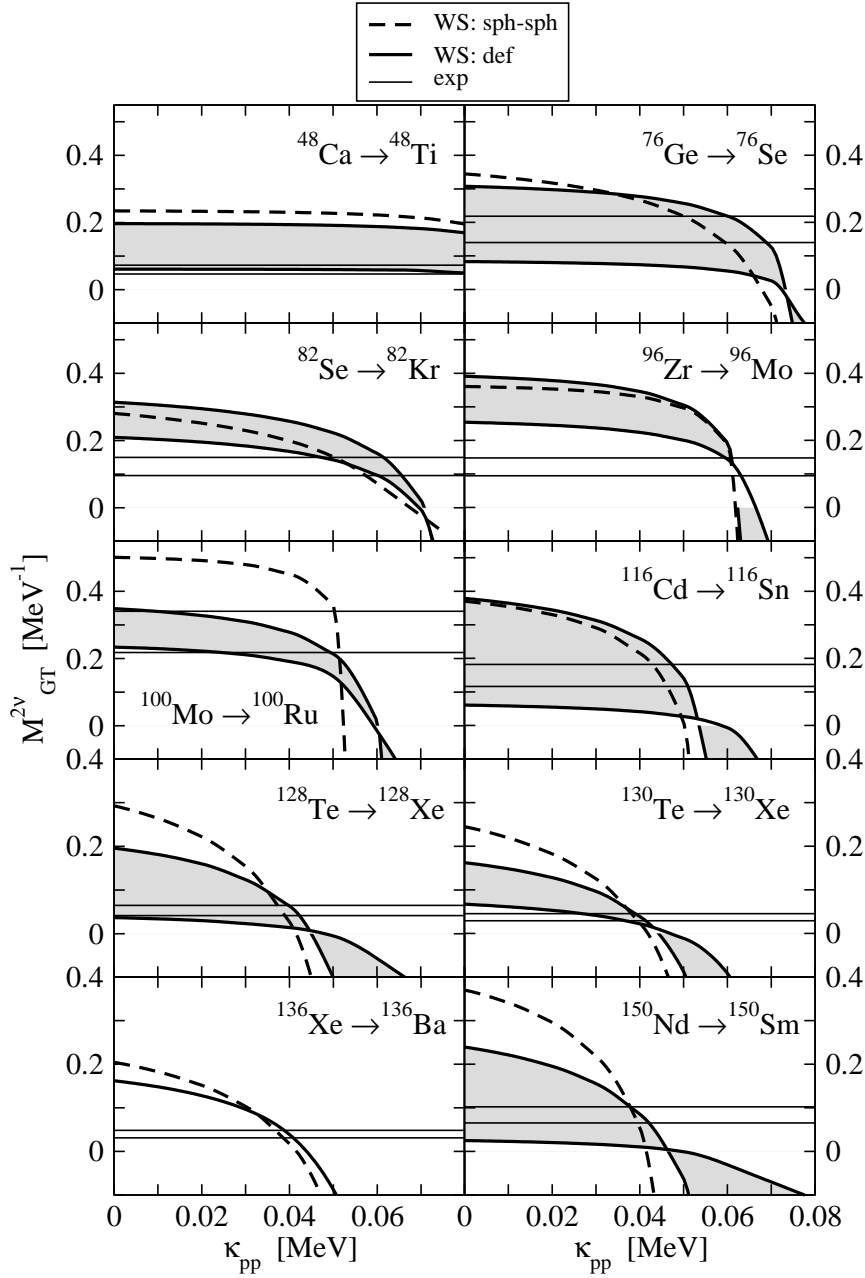,width=0.8\textwidth}
\caption{$2\nu\beta\beta-$decay matrix elements calculated with
Woods-Saxon potentials as a function of the 
particle-particle interaction strength $\kappa ^{pp}_{GT}$.
Dashed lines correspond to the results assuming spherical nuclei.
Solid lines correspond to the results obtained by using the maximum
and minimum differences between the experimental deformations of parent
and daughter (see Table 2). Horizontal lines are the experimental
$M^{2\nu}_{\rm GT}$ extracted from Ref. \protect\cite{barabash} using 
$g_A=1.0$ and $g_A=1.25$.}
\end{figure}

\newpage

\begin{figure}[t]
\psfig{file=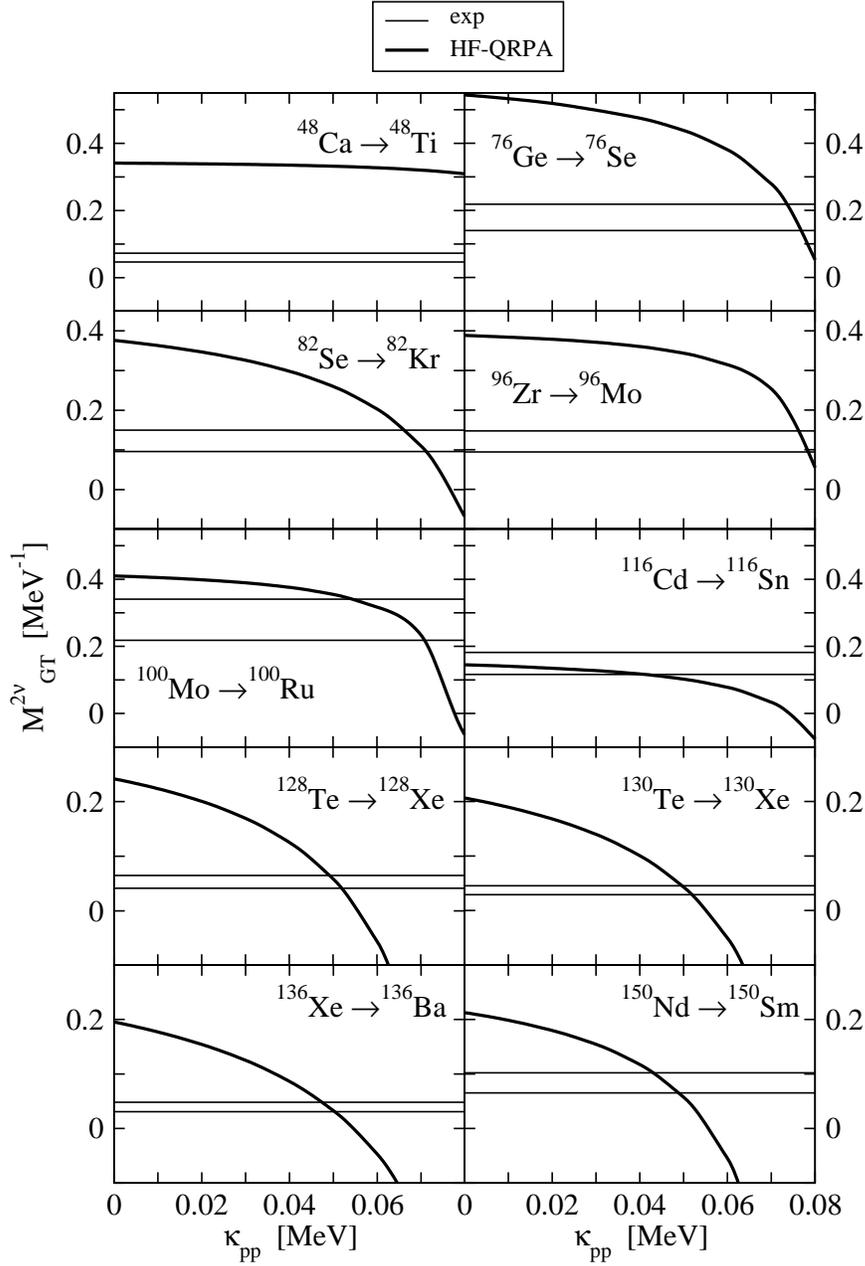,width=0.8\textwidth}
\caption{$2\nu\beta\beta-$decay matrix elements obtained from
Skyrme(Sk3) deformed Hartree-Fock calculations as a function of the 
particle-particle interaction strength $\kappa ^{pp}_{GT}$.
Horizontal lines are the experimental
$M^{2\nu}_{\rm GT}$ extracted from Ref. \protect\cite{barabash} using 
$g_A=1.0$ and $g_A=1.25$.}
\end{figure}

\newpage

\begin{figure}[t]
\psfig{file=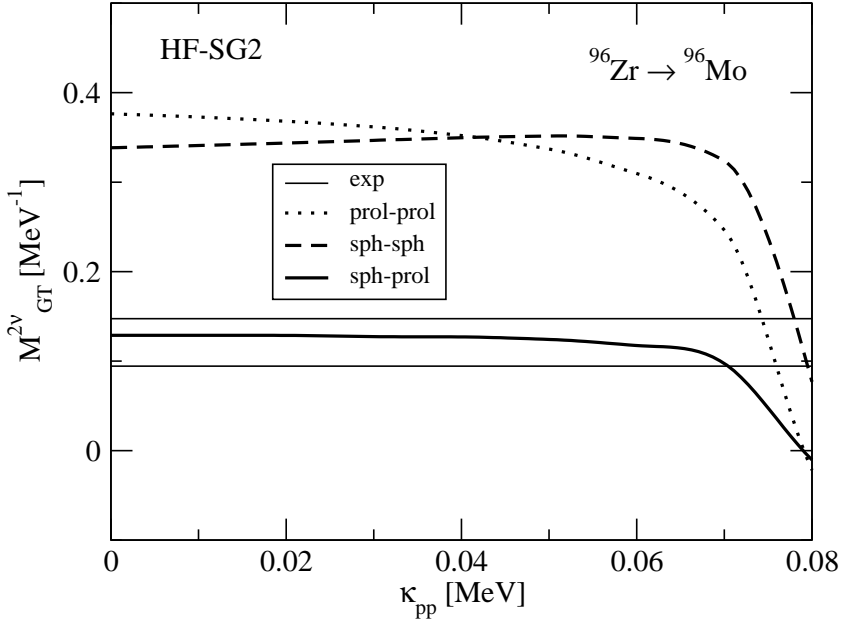,width=0.8\textwidth}
\caption{$2\nu\beta\beta-$decay matrix elements of $^{96}$Zr obtained from
Skyrme(SG2) deformed Hartree-Fock calculations as a function of the 
particle-particle interaction strength $\kappa ^{pp}_{GT}$.
The dotted curve corresponds to calculations using the prolate shapes
for parent and daughter $\beta_{\rm p}=0.147,\,\beta_{\rm d}=0.119$ 
(see Table 2). The dashed curve is for spherical shapes and the solid curve 
is for spherical parent and prolate daughter 
$\beta_{\rm p}=0.016,\,\beta_{\rm d}=0.119$ (see Table 2).
Horizontal lines are the experimental
$M^{2\nu}_{\rm GT}$ extracted from Ref. \protect\cite{barabash} using 
$g_A=1.0$ and $g_A=1.25$.}
\end{figure}
\end{center}
\end{document}